\documentclass[conference]{IEEEtran}
\IEEEoverridecommandlockouts
\usepackage{cite}
\usepackage{amsmath,amssymb,amsfonts}
\usepackage{algorithmic}
\usepackage{graphicx}
\usepackage{textcomp}
\usepackage{lmodern} % Load modern fonts to avoid font shape issues
\usepackage{xcolor}
\usepackage{amsmath,graphicx}
\usepackage{cite}
\usepackage{subfigure}
\usepackage{graphicx}
\usepackage{subcaption}
\usepackage[linesnumbered, ruled]{algorithm2e} % Retain only one algorithm package
\usepackage{enumitem}
\usepackage{tikz}
\usepackage[caption=false]{subfig} % 避免与caption宏包冲突，caption=false选项有时可避免一些问题
\usepackage{subcaption}
\usetikzlibrary{arrows.meta, positioning}

\def\BibTeX{{\rm B\kern-.05em{\sc i\kern-.025em b}\kern-.08em
    T\kern-.1667em\lower.7ex\hbox{E}\kern-.125emX}}
\begin{document}
\captionsetup[figure]{name={Fig.},labelsep=period}
\title{Zeroth-Order Blind Interference Suppression for Multi-RIS-Aided Wireless Systems
}  
\author{ Binyao Ma$^{1}$, Peilan Wang$^{ 1}$, Bin Wang$^{2}$, and Jun Fang$^{1}$, \textit{Senior Member, IEEE}\\
\thanks{This work was supported in part by the National Science Foundation of China under Grant 62501121, and in part by the National Key Laboratory of Wireless Communications Foundation under Grant IFN202409.}%
		$^{1}$ University of Electronic Science and Technology of China, Chengdu, China, 611731\\
        $^{2}$ Shenzhen Research Institute of Big Data, Shenzhen, Guangdong, China, 518172}
        % Emails:  binyao.ma@std.uestc.edu.cn; peilan\_wangle@uestc.edu.cn;15129884427@163.com;JunFang@uestc.edu.cn
	
    % {Peilan Wang$^{ 1}$, Binyao Ma$^{1}$, Jun Fang$^{1}$, Bin Wang$^{2}$, and Hongbin Li$^3$ \\
    % $^{1}$ University of Electronic Science and Technology of China, Chengdu, China, 611731\\
    % $^{2}$ National University of Defense Technology, Xian, China, 710106\\
    % $^{3}$ Stevens Institute of Technology, Hoboken, New Jersey, USA, 07030\\
    % Emails: peilan\_wangle@uestc.edu.cn;  binyao.ma@std.uestc.edu.cn; JunFang@uestc.edu.cn; \\
    % 15129884427@163.com; Hongbin.Li@stevens.edu

\maketitle

\begin{abstract}
    % In this paper, we consider a multi-reconfigurable intelligent surface (RIS)-aided single-input single-output (SISO) system in the presence of several unknown strong interference signals. We aim to maximize the system signal-to-interference-plus-noise ratio (SINR) without requiring explicit interference channel state information. Specifically, we formulate a black-box optimization problem over RIS reflection phase shifts based solely on received power measurements. To address this challenge, we propose a group-based zeroth-order adaptive moment method (ZO-AdaMM) in which adjacent reflecting elements share a common phase shift, effectively reducing the optimization dimensionality. The proposed method is further extended to accommodate practical low-resolution RIS phase shifters with discrete quantization constraints. Simulation results demonstrate that the proposed approach achieves superior performance compared to state-of-the-art methods in terms of both SINR attainment and sample complexity, making it particularly suitable for resource-constrained multi-RIS deployments. 

    In this paper, we study measurement-driven signal-to-interference-plus-noise ratio (SINR) maximization for a multi-reconfigurable intelligent surface (RIS)-aided single-input single-output (SISO) system with unknown strong interference sources. Specifically, the objective is to optimize the reflection coefficients such that the SINR is maximized at the receiver. As the interference channels are unknown, such an optimization problem is a black-box optimization problem with an objective function whose closed-form analytical expression is unknown. To address the high-dimensional black-box optimization problem with discrete variable constraints, we introduce a group-based phase parameterization that significantly reduces the search dimension. Building on this model, we develop a group-based zeroth-order adaptive moment (ZO-AdaMM) algorithm. Simulation results show that the proposed grouping strategy markedly accelerates the convergence speed and achieves a superior interference suppression performance under limited measurement budgets, especially in the small-budget regime.

\end{abstract}
% sdf\cite{yan2025power}
%
\section{Introduction}
Reconfigurable intelligent surface (RIS) has been a paradigm shift for wireless systems \cite{11353466,wu2024intelligent} due to their capability to reshape wireless propagation through programmable phase shifts, thereby enabling capacity/coverage enhancement\cite{wang2020intelligent}, interference mitigation\cite{wang2025creating}, and energy-efficient transmission\cite{huang2019reconfigurable}. Deploying multiple RISs is essential to improve system throughput and robustness by providing flexible and controllable propagation paths \cite{basar2024reconfigurable}. However, acquiring channel state information (CSI) is inherently challenging in RIS-assisted systems\cite{zhao2025reconfigurable}, and this difficulty is further exacerbated by the increasing number of RISs\cite{liang2025survey}. This is due to the multiple reflection links introduced by two or more RISs, especially in the presence of non-cooperative interferences.

% , but also enlarges the control dimension and makes channel acquisition extremely challenging: the number of cascaded channels grows rapidly with the number of RISs, and bottlenecked by channel state information (CSI) acquisition\cite{zheng2022survey}, thus CSI-based optimization incurs substantial training and computation overhead\cite{yao2023blind}.

% Blind/measurement-driven optimization emerges as a solution, updating RIS configurations via observable metrics (e.g., received power) without CSI\cite{yao2023blind}. 

Recently, blind beamforming without explicit CSI acquisition has emerged as a promising solution \cite{yao2023blind,wang2025low,10959084,11310555}. In this approach, the reflection coefficients of the RIS are configured by collecting a large number of received signal power measurements and optimizing each coefficient based on maximizing the conditional expectation. Specifically, the authors in \cite{yao2023blind} developed multi-RIS sequential conditional sample mean (SCSM) method to estimate the conditional mean for each discrete phase shift state and update phase sequentially. SCSM can achieve a quartic $\mathcal{O}(M^4)$ signal-to-noise ratio (SNR) gain in a double-RIS system under favorable rank-one/line-of-sight (LoS) conditions and its benefits have been validated in prototypes, where $M$ is the number of reflected elements. Nevertheless, CSM-based methods rely on a large number of samples to approximate the conditional expectation, which can lead to a prohibitively high sample complexity.

To efficiently leverage the inherent structure of the received signal power, the work in \cite{10959084,wang2025creating} address the problem of blind interference suppression by proposing to implicitly infer the null space of the interference covariance matrix based solely on received signal power measurements. Although achieving superior performance, the proposed method requires control of the reflection amplitudes as well as the phase shifts, both of which are assumed to be continuous variables, and performance degradation may occur otherwise. To alleviate the hardware burden of RIS with phase-shift-only control, we introduced a derivative-free approach in \cite{11310555} that treats blind interference suppression as a black-box problem \cite{munoz2015algorithm} and relies solely on received signal power measurements to construct an objective function that maximizes the signal-to-interference-plus-noise ratio (SINR).

In this paper, we consider a more challenging blind interference suppression scenario involving multi-RIS with discrete phase shifts only. To further reduce hardware complexity and accelerate convergence, we propose a group-based zeroth-order adaptive moment method (ZO-AdaMM). By grouping adjacent elements to reduce the effective dimension, the proposed method achieves faster, more sample-efficient convergence and higher SINR in the small-budget regime, as revealed by simulation results.

\section{System Model}
\begin{figure}[htbp]
    \centering
    \includegraphics[width=0.75\linewidth,trim=0 0 0 0.4cm,clip]{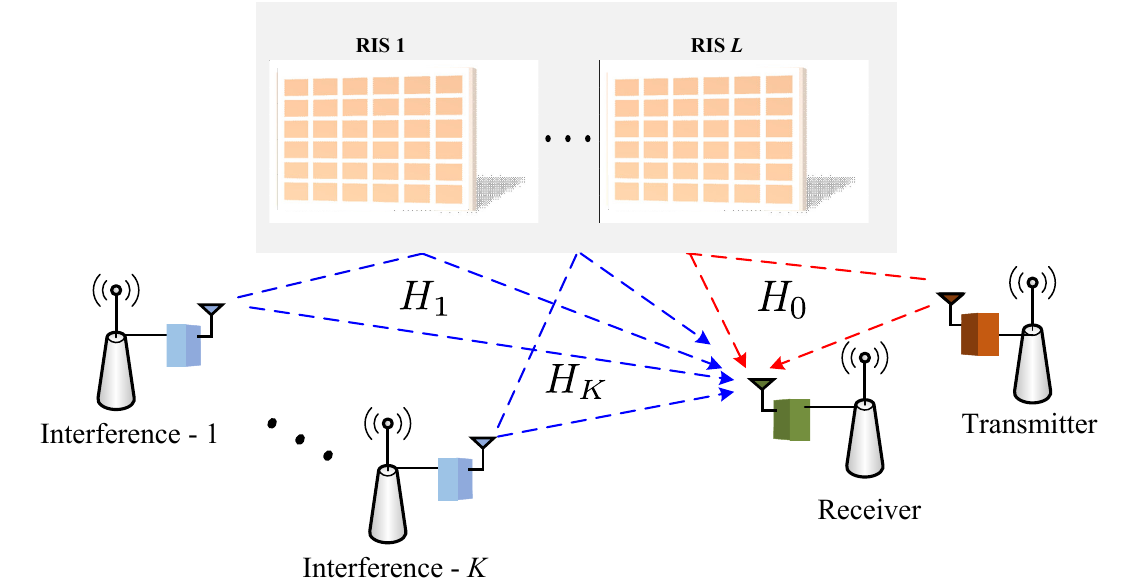} 
    \caption{The RISs-assisted wireless communication system.}
    \label{fig:double_RIS_model}
\end{figure}
As shown in Fig.\ref{fig:double_RIS_model}, this paper considers a multi-RIS-assisted SISO system in the presence of $K$ unknown strong interference signals. By leveraging the channel reconfiguration capabilities of RISs, we aim to suppress the unknown interference signals while simultaneously enhancing legitimate communication performance.

The $l$th RIS employs a uniform planar array (UPA) structure with $M_l=M_{l,x} \times M_{l,y}$ reflecting elements, and its reflection phase shift matrix is given by
\begin{equation}
\boldsymbol{\Theta}_{l}  \triangleq \mathrm{diag} (e^{j\boldsymbol{\theta}_l^H})
% \mathrm{diag}\!\left(e^{j\theta_{l,1}},\ldots,e^{j\theta_{l,M_l}}\right),
\end{equation}
where $\boldsymbol{\theta}_l=[\theta_{l,1} \phantom{0} \ldots \phantom{0} \theta_{l,M_l}]^H$ denotes the reflection phase shift vector, and  $\theta_{l,m}$ denotes the phase shift of the $m$th element at the $l$th RIS. In practice, the phase shift often is selected from a finite set of discrete values due to hardware limitations. With $B$ quantization bits, each phase value can be represented by one of $2^{B}$ discrete states, i.e., there are $|\mathcal{F}|=2^{B}$ admissible phase levels with quantization step defined as $\Delta \triangleq \frac{2\pi}{2^{B}}$.
% Accordingly, the discrete set is defined as
With B quantization bits, each phase shift $\theta_{l,m}$ can be selected from a set of discrete states; i.e., there are
\begin{equation}
\mathcal{F} \triangleq \{0,\Delta,2\Delta,\ldots,(2^{B}-1)\Delta\}.
\end{equation}
By stacking all $L$ reflecting phase shift vectors, the aggregated phase-shift vector of all RIS elements is defined as
\begin{equation}
% \boldsymbol{\theta}_l \triangleq 
% [\theta_{1,1},\ldots,\theta_{1,M_1},\ldots,\theta_{L,1},\ldots,\theta_{L,M_L}]^{T}.
\boldsymbol{\theta}\triangleq [\boldsymbol{\theta}_1^H \phantom{0} \ldots  \phantom{0}  \boldsymbol{\theta}_L^H]^H
% = [\boldsymbol{\theta}_l,\ldots,\theta_{1,M_1},\ldots,\theta_{L,1},\ldots,\theta_{L,M_L}]^{T}.
\label{eqn:theta}
\end{equation}
In general, the composite channel from the transmitter to the receiver comprises the direct link, the reflected links via a single RIS, and potential multi-hop reflected links via multi-RIS, as illustrated in Fig. 1. To sidestep the intricate channel modeling associated with RISs, we define
$H_k(\boldsymbol{\theta}), k=0, 1, \ldots, K$
as the effective end-to-end channel from the legitimate transmitter ($k=0$) and the $k$th interference source ($k=1,2,\ldots, K$) to the receiver.

% By jointly accounting for the direct link, single-bounce RIS reflections, and possible multi-hop reflections among RISs, the equivalent cascaded channel from the transmitter to the receiver can be expressed as a function of $\boldsymbol{\theta}$:
% \begin{equation}
% H_0(\boldsymbol{\theta}) \in \mathbb{C},
% \end{equation}
% and the equivalent cascaded channel from the $k$th interferer to the receiver is denoted by
% \begin{equation}
% H_k(\boldsymbol{\theta}) \in \mathbb{C}, \quad k=1,\ldots,K.
% \end{equation}
% the cascaded channels $H_0$ and $\{H_k\}_{k=0}^{K}$ between the legitimate transmitter/the interferers and the receiver  are not identical.

% Then the received signal at the receiver is

Clearly, the received signal at the receiver is the superposition of the legitimate signal and interference signals, i.e.,
\begin{equation}
y(t) = H_0(\boldsymbol{\theta}) s(t) + \sum\nolimits_{k=1}^{K} H_k(\boldsymbol{\theta}) j_k(t) + \epsilon(t),
\label{eq:rx_signal_general}
\end{equation}
in which $s(t)\sim \mathcal{CN}(0,\sigma_0^2)$ is the transmitted signal from the legitimate transmitter, $j_k(t)\sim \mathcal{CN}(0,\sigma_k^2)$ denotes the interference signal from the $k$th interferer, and $\epsilon(t)\sim\mathcal{CN}(0,\sigma_w^2)$ is additive white Gaussian noise.

In this paper, we consider block fading channels where the channel remains constant over a given coherence time block. In addition, the desired signal and the interference signals are assumed to be mutually uncorrelated.

Since the interference signals and noise are unknown at the receiver, we obtain quadratic measurements by averaging over sufficiently many samples to capture their second-order statistics. Specifically, at the receiver, only the average received signal power can be observed, which is given by
\begin{align}
	P(\boldsymbol{\theta}) &\triangleq \mathbb{E}[|y(t)|^2]\\
	&= \underbrace{\mathbb{E}[|H_0(\boldsymbol{\theta})s(t)|^2]}_{P_S(\boldsymbol{\theta})}
	+ \underbrace{\mathbb{E}\!\left[\left|\sum\nolimits_{k=1}^K H_k(\boldsymbol{\theta})j_k(t)\right|^2\right]}_{P_I(\boldsymbol{\theta})}
	+ \underbrace{\sigma_w^2}_{P_{N}} \nonumber\\
    &= \sigma_0^2 \mathbb{E}[|H_0(\boldsymbol{\theta})|^2]
    + \sum\nolimits_{k=1}^{K} \sigma_k^2 \mathbb{E}[|H_k(\boldsymbol{\theta})|^2]
    + \sigma_w^2 .
    \label{eq:power_general}
\end{align}

The received SINR directly using only received power observations, which is defined as

\begin{equation}
\mathrm{SINR}(\boldsymbol{\theta}) \triangleq
\frac{P_S(\boldsymbol{\theta})}
{P_I(\boldsymbol{\theta}) + \sigma_w^2}.
\label{eq:sinr_general}
\end{equation}
% The interference sources are usually non-cooperative, and thus their CSI, denoted as $\boldsymbol{H}_k(\cdot), k=1,2,...,K$ is typically unavailable. This makes direct estimation of the interference power $P_I(\boldsymbol{\theta})$ impractical. 
For legitimate transmission, the desired signal power $P_S(\boldsymbol{\theta})$ can be calculated using effective channel function $H_0(\cdot)$, which is acquired via pilot-aided estimation methods \cite{wang2025low,10959084}. Since the total received power $P(\boldsymbol{\theta})$ can be directly measured and $P_S(\boldsymbol{\theta})$ is known, the SINR in \eqref{eq:sinr_general} can be equivalently evaluated as
\begin{equation}
\mathrm{SINR}(\boldsymbol{\theta}) =
\frac{P_S(\boldsymbol{\theta})}
{P(\boldsymbol{\theta}) - P_S(\boldsymbol{\theta})}.
\label{eq:sinr_general_2}
\end{equation}

\section{Problem Formulation}
In this paper, the objective is to maximize the received SINR by optimizing the RIS phase shift vector without any knowledge of the CSI relevant to interference sources. The optimization problem is formulated as
\begin{align}
(\mathrm{P1})~~
\max_{\boldsymbol{\theta}}~~ & \mathrm{SINR}(\boldsymbol{\theta})\\
\mathrm{s.t.}~~ & \theta_{l,m} \in \mathcal{F}, \quad \forall l,m,
\end{align}
where $\mathrm{SINR}(\boldsymbol{\theta})$ is defined in \eqref{eq:sinr_general}. 

Problem (P1) is non-convex and combinatorial due to the fractional objective and the discrete constraints on the RIS reflection coefficients. When the CSI of both the desired and interference links is available, such a problem can be solved by first obtaining a continuous solution via manifold optimization\cite{boumal2014manopt} methods and then projecting it onto the discrete set. However, the interference CSI is typically unavailable as the interference sources are inherently non-cooperative. Consequently, the resulting SINR values we can only obtain via measurements as defined in \eqref{eq:sinr_general_2}. This renders problem (P1) a black-box optimization problem\cite{munoz2015algorithm}, for which traditional CSI-dependent methods are inapplicable.

% ===

% Problem (P1) is non-convex due to the fractional quadratic form and the unit-modulus constraints on the RIS reflection coefficients.

% A major challenge in solving the quadratic programming problem (P1) is that the interference power $P_I(\boldsymbol{\theta})$ is unknown to the receiver. Although (P1) could be addressed if full CSI were available, acquiring such information is impractical in the presence of multiple unknown interferers. 

% In the following sections, we develop a zeroth order optimization framework that directly maximizes the SINR using only received power observations, without explicitly estimating $P_I(\boldsymbol{\theta})$.

\section{Proposed Method}
% The reflection coefficients of the RISs are characterized by the equation(\ref{eqn:theta}) where each phase variable $\theta_{l,i}$ corresponds to an individual reflecting element on the RIS. Although such an element-wise modeling provides the maximum degree of freedom for phase control, the optimization dimension grows linearly
% with the number of reflecting elements. As a result, both the algorithmic
% complexity and the measurement overhead increase rapidly when large-scale RISs
% are employed.

% To address this issue, this paper introduces a group-based RIS phase modeling
% approach. Without altering the underlying channel model or the mathematical form of the received signal expression, this method enforces phase sharing among groups of adjacent reflecting elements, thereby effectively reducing the optimization dimension.

In this section, we present a group-based zeroth-order (ZO) passive beamforming scheme. To reduce the optimization dimension, the RIS elements are partitioned into sub-groups of adjacent reflecting elements, where elements within each group share a common phase shift. We first develop the algorithm under a continuous phase-shift assumption and subsequently extend it to discrete settings. By leveraging a ZO gradient approximation to determine the search direction, the proposed algorithm iteratively updates the group phase shifts based on SINR measurements. For the discrete case, the updated variables are projected onto a feasible discrete set to comply with practical hardware constraints.

\subsection{Grouped Reflection Phase Shifts Modeling}
\begin{figure}[htbp]
    \centering
    \includegraphics[width=0.8\linewidth,trim=0 0 0 0.15cm,clip]{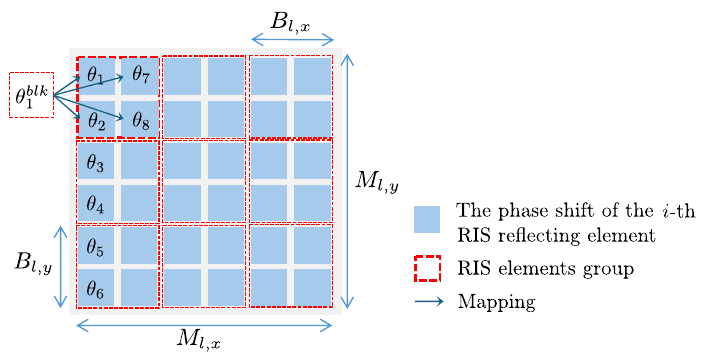}
    \caption{ Grouped reflection phase shifts model.}
    \label{fig:IRS_group}
\end{figure}

As shown in Fig. \ref{fig:IRS_group}, each RIS is partitioned into $N_{x,l}\times N_{y,l}$ non-overlapping rectangular groups, each with a size of $B_x \times B_y$, where
% . Accordingly, the number of groups along the horizontal and vertical directions is given by
\begin{align}
N_{x,l} = \left\lceil \tfrac{M_{l,x}}{B_x} \right\rceil, 
N_{y,l} = \left\lceil \tfrac{M_{l,y}}{B_y} \right\rceil.
\end{align}
Let $\boldsymbol{\theta}^{blk} =[\theta_1^{blk} \phantom{0} \ldots \phantom{0} \theta_d^{blk}]^T\in \mathbb C^{D_l}, D_l=N_{x,l}N_{y,l}$ denote the group phase shift vector comprising  common phase shifts among adjacent reflecting elements. The overall phase shift vector $\boldsymbol{\theta}$ can then be recast as
\begin{equation}
\boldsymbol{\theta} = \mathcal{M}(\boldsymbol{\theta}^{\mathrm{blk}}),
\end{equation}
where $\mathcal{M}(\cdot)$ is a deterministic one-to-many mapping.

\subsection{Group-Based ZO Optimization}

To begin with, we 
%relax the discrete constraints in (P1) by allowing $\theta_{l,m} \in [0,2\pi)$ and subsequently 
propose a \textbf{ZO-AdaMM-based algorithm} to solve the resulting continuous black-box problem. By employing the group-based modeling, the effective optimization dimension is  reduced to
$D=\sum\nolimits_{l=1}^{L} D_l$. 

At the $n$-th iteration, a random direction $\boldsymbol{u}[n]\in\mathbb{R}^{D}$ is generated, and a central-difference ZO gradient estimator is constructed as
\begin{align}
\widehat{\nabla} f(\boldsymbol{\theta}^{\mathrm{blk}}[n])=\frac{D}{2\mu}
\Big[
f(\boldsymbol{\theta}_{+}^{\mathrm{blk}}[n])
-
f(\boldsymbol{\theta}_{-}^{\mathrm{blk}}[n])
\Big]\boldsymbol{u}[n],
\label{eq:18}
\end{align}
where $\mu>0$ is a smoothing parameter, $\boldsymbol{\theta}_{\pm}^{\mathrm{blk}}[n] \triangleq \boldsymbol{\theta}^{\mathrm{blk}}[n]\pm\mu\boldsymbol{u}[n],$ and $f(\boldsymbol{\theta}^{\mathrm{blk}})\triangleq\mathrm{SINR}(\mathcal{M}(\boldsymbol{\theta}^{\mathrm{blk}}))$ denotes the measured SINR value when RISs are configured with $\boldsymbol{\theta} = \mathcal{M}(\boldsymbol{\theta}^{blk})$. The ZO-AdaMM update at the $n$-th iteration is then carried out as
\begin{align}
\widehat{\nabla} g[n]  &\triangleq\widehat{\nabla} f(\boldsymbol{\theta}^{\mathrm{blk}}[n]),\\
\boldsymbol{m}[n]
&= \beta_1 \boldsymbol{m}[n-1]
+ (1-\beta_1)\widehat{\nabla} g[n],\\
\boldsymbol{v}[n]
&= \beta_2 \boldsymbol{v}[n-1]
+ (1-\beta_2)\widehat{\nabla} g[n]
\odot \widehat{\nabla} g[n],\\
\hat{\boldsymbol{v}}[n]
&= \max(\hat{\boldsymbol{v}}[n-1], \boldsymbol{v}[n]),
\end{align}
and the group phase shifts vector is updated according to
\begin{align}
\boldsymbol{\theta}[n+1]^{\mathrm{blk}}
= \boldsymbol{\theta}^{\mathrm{blk}}[n]
+ \alpha[n] \hat{\boldsymbol{V}}[n]^{-1/2} \boldsymbol{m}[n],
\end{align}
where $\hat{\boldsymbol{V}}[n]=\mathrm{diag}(\hat{\boldsymbol{v}}[n])$ and $\alpha[n]$ denotes the step size. The proposed grouped ZO method is guaranteed to converge to a critical point\cite{chen2019zo}.

% Through the above design, the proposed ZO-AdaMM framework enables efficient and stable RIS phase self-optimization under unknown interference statistics, limited measurement budgets, and large-scale RIS deployments, while preserving the original system model and received signal formulation.

% \begin{algorithm}[htbp]
% \caption{The proposed Group-Based ZO-AdaMM for solving (P1)}
% \label{alg:ZO-AdaMM}
% \KwIn{
% Initial group phase vector $\boldsymbol{\theta}^{\mathrm{blk}}[0]$, step size $\alpha$, smoothing parameter $\mu$, momentum parameters $\beta_1,\beta_2$
% }
% \KwOut{Optimized phase shifts vector $\boldsymbol{\theta}$}

% Initialize $\boldsymbol{m}_0=\boldsymbol{0}$, $\boldsymbol{v}_0=\hat{\boldsymbol{v}}_0=\boldsymbol{0}$\;

% \For{$n=1,2,\ldots,N-1$}{

% % Update first moment:
% $\widehat{\nabla} g[n]  \triangleq\widehat{\nabla} f(\boldsymbol{\theta}^{\mathrm{blk}}[n])$

% $\boldsymbol{m}[n]=\beta_1\boldsymbol{m}[n-1]+(1-\beta_1)\widehat{\nabla} g[n]$\;

% % Update second moment:
% $\boldsymbol{v}[n]=\beta_2\boldsymbol{v}[n-1]+(1-\beta_2)\widehat{\nabla} g[n]\odot\widehat{\nabla} g[n]$\;

% % Update maximum second moment:
% $\hat{\boldsymbol{v}}[n]=\max(\hat{\boldsymbol{v}}[n-1],\boldsymbol{v}[n])$\;

% % Update group phase vector:
% $\boldsymbol{\theta}[n+1]^{\mathrm{blk}}=\boldsymbol{\theta}^{\mathrm{blk}}[n]+\alpha\hat{\boldsymbol{V}}[n]^{-1/2}\boldsymbol{m}[n]$\;
% }
% \Return{$\boldsymbol{\theta}^{\mathrm{blk}}[N],\boldsymbol{\theta}[N]=\mathcal{M}(\boldsymbol{\theta}^{\mathrm{blk}}[N])$}
% \end{algorithm}

\begin{figure*}[!t]
    \centering
    \begin{minipage}{0.32\textwidth}
        \vspace{0pt}
        \centering
        \includegraphics[height=4.2cm,keepaspectratio]{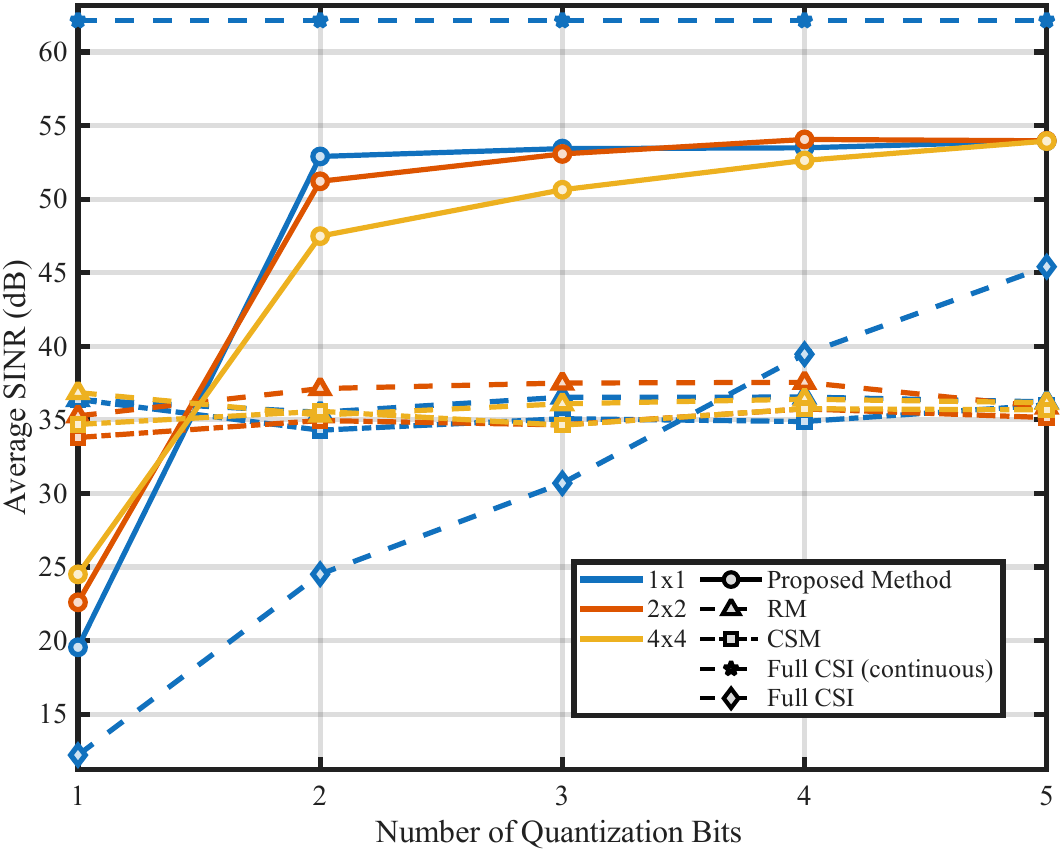}
        \caption{SINR vs. the number of quantization bits with $T=2000$.}
        \label{bit_sweep}
    \end{minipage}
    \hfill
    \begin{minipage}{0.33\textwidth}
        \vspace{0pt}
        \centering
        \includegraphics[height=4.2cm,keepaspectratio]{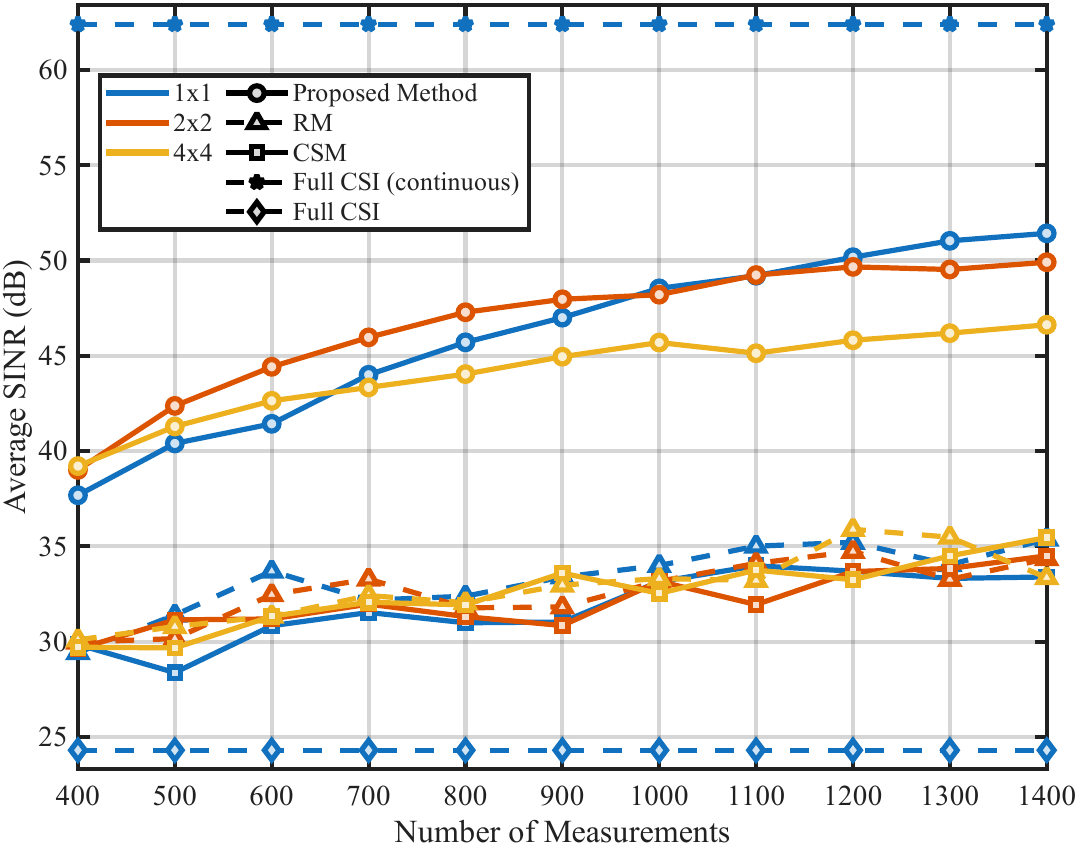}
        \caption{SINR vs. the number of measurements $T$ for different group sizes.}
        \label{meas_sweep}
    \end{minipage}
    \hfill
    \begin{minipage}{0.32\textwidth}
        \vspace{0pt}
        \centering
        \includegraphics[height=4.2cm,keepaspectratio]{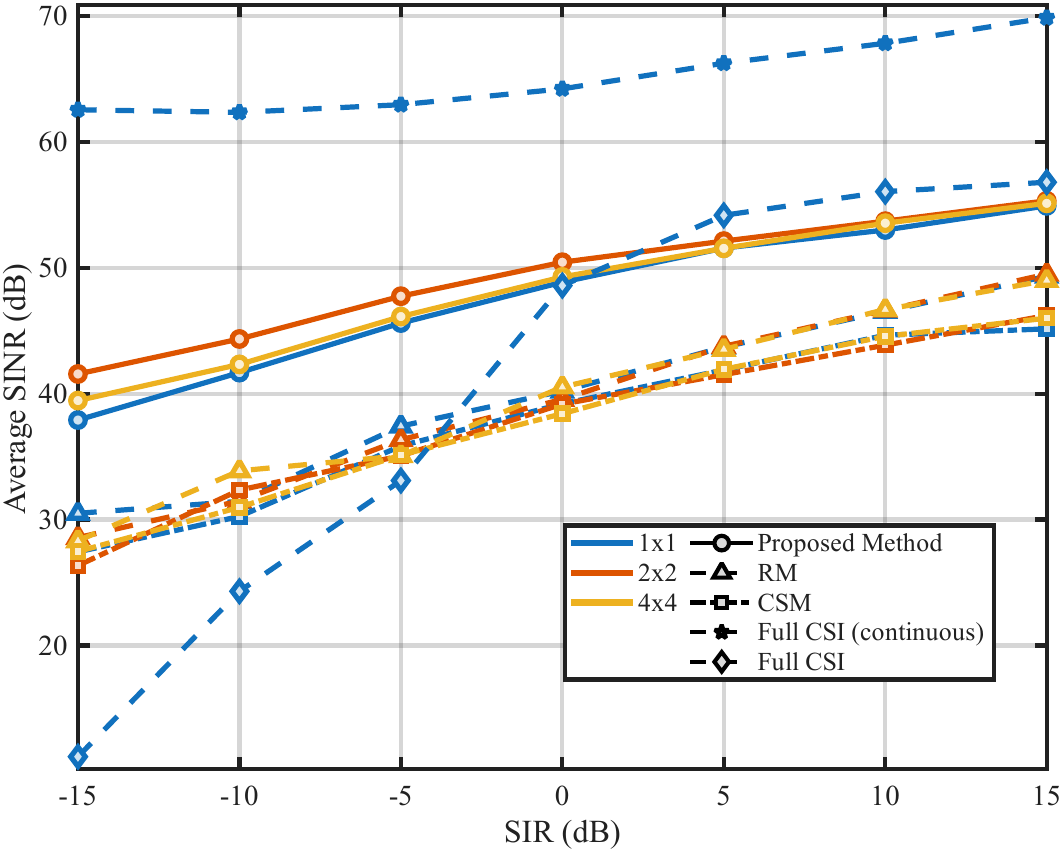}
        \caption{SINR vs. SIR with a fixed $T=600$.}
        \label{sir_sweep}
    \end{minipage}
\end{figure*}
\subsection{Extension to Discrete Phase Shifts}
To enforce the discrete constraint, we define an element-wise quantization (projection) operator $\mathcal{Q}(\cdot)$ as
\begin{equation}
\mathcal{Q}(x)\triangleq \arg\min_{\phi\in\mathcal{F}}|x-\phi|, x\in[0,2\pi).
\end{equation}

At the $n$-th iteration, the two probing points used in the central-difference estimator in \eqref{eq:18} are quantized as
\begin{align}
\tilde{\boldsymbol{\theta}}_{\pm}^{\mathrm{blk}}[n] &= \mathcal{Q}\!\left(\boldsymbol{\theta}^{\mathrm{blk}}[n]\pm\mu\boldsymbol{u}[n]\right).
\end{align}
% which guarantees that all tested phase configurations satisfy the hardware limitation. 
The ZO gradient estimator is computed by replacing the two function evaluations in \eqref{eq:18} with $f(\tilde{\boldsymbol{\theta}}_{+}^{\mathrm{blk}}[n])$ and $f(\tilde{\boldsymbol{\theta}}_{-}^{\mathrm{blk}}[n])$. After each ZO-AdaMM update, the resulting phase vector is projected back onto the feasible set:
\begin{equation}
    \tilde{\boldsymbol{\theta}}^{\mathrm{blk}}[n+1]\leftarrow \mathcal{Q}(\boldsymbol{\theta}^{\mathrm{blk}}[n+1]).
\end{equation}
Notably, quantization may compromise the convergence guarantees of the ZO-AdaMM algorithm. Nevertheless, the simple projection approach still exhibits satisfactory performance, as demonstrated in the following numerical simulation.

\begin{algorithm}[htbp]
\caption{Group-Based ZO-AdaMM}
\label{alg:DZO-AdaMM}
\KwIn{
Quantized group phase vector $\tilde{\boldsymbol{\theta}}^{\mathrm{blk}}[0]$,
step size $\alpha$,
smoothing parameter $\mu$,
momentum parameters $\beta_1,\beta_2$,
discrete set $\mathcal{F}$.
}
\KwOut{Optimized discrete phase shifts vector $\tilde{\boldsymbol{\theta}}$.}

\BlankLine

Initialize $\boldsymbol{m}[0]=\boldsymbol{0}$, $\boldsymbol{v}[0]=\hat{\boldsymbol{v}}[0]=\boldsymbol{0}$\;

\For{$n=1,2,\ldots,N-1$}{

$\tilde{\boldsymbol{\theta}}_{\pm}^{\mathrm{blk}}[n] = \mathcal{Q}\!\left(\tilde{\boldsymbol{\theta}}^{\mathrm{blk}}[n]\pm\mu\boldsymbol{u}[n]\right)$\;

$\widehat{\nabla} g[n]= \dfrac{D}{2\mu}
\Big(f(\tilde{\boldsymbol{\theta}}_{+}^{\mathrm{blk}}[n])-
f(\tilde{\boldsymbol{\theta}}_{-}^{\mathrm{blk}}[n])\Big)\boldsymbol{u}[n]$\;

$\boldsymbol{m}[n+1]= \beta_1\boldsymbol{m}[n]+(1-\beta_1)\widehat{\nabla} g[n]$\;
$\boldsymbol{v}[n+1]= \beta_2\boldsymbol{v}[n]+(1-\beta_2)\widehat{\nabla} g[n]\odot\widehat{\nabla} g[n]$\;
$\hat{\boldsymbol{v}}[n+1]= \max\!\big(\hat{\boldsymbol{v}}[n],\boldsymbol{v}[n+1]\big)$\;
$\boldsymbol{\theta}^{\mathrm{blk}}[n+1]= \tilde{\boldsymbol{\theta}}^{\mathrm{blk}}[n]
+\alpha \hat{\boldsymbol{V}}[n+1]^{-1/2}\boldsymbol{m}[n+1]$\;

$\tilde{\boldsymbol{\theta}}^{\mathrm{blk}}[n+1]= \mathcal{Q}\!\left(\boldsymbol{\theta}^{\mathrm{blk}}[n+1]\right)$\;

}

\Return{$\tilde{\boldsymbol{\theta}}^{\mathrm{blk}}[N],\ \tilde{\boldsymbol{\theta}}[N]=\mathcal{M}(\tilde{\boldsymbol{\theta}}^{\mathrm{blk}}[N])$}\;
\end{algorithm}

\section{Simulation Results}
In this section, we evaluate the performance of the proposed group-based ZO-AdaMM method for a double-RIS-assisted communication system \cite{yan2025power}. Each RIS comprises $16\times16$ reflecting elements ($M_1= M_2= 256$), with each element employing a low-resolution 2-bit phase shifter.
% All simulations are conducted under practical hardware constraints, where each RIS is equipped with a 2-bit phase shifter and consists of $16\times16$ reflecting elements, i.e., $M_1=M_2=256$.

The transmitter, receiver, RIS~1, and RIS~2 are located at $(1,0,2)$, $(1,50,0)$, $(0,-5,1)$, and $(0,55,1)$ meters, respectively. In this work, we focus on the case of a single interference source, i.e., $K=1$, which is located at $(55,0,0)$ meters. Large-scale fading is modeled using distance-dependent path-loss functions with different parameters. Specifically, the path loss for the direct link is $32.6 + 36.7\log_{10}d$ , while the corresponding path loss for the reflected link is $30 + 22\log_{10}d$, where $d$ is the distance in meters. The receiver noise power is set to as $\sigma_w^2 \approx -107$ dBm. The signal-to-interference ratio (SIR) is defined as $\mathrm{SIR} \triangleq \sigma_0^2 /\sum\nolimits_{k=1}^{K} \sigma_k^2$, 
% \begin{equation}
% \mathrm{SIR} \triangleq \frac{\sigma_0^2 }{\sum_{k=1}^{K} \sigma_k^2 },
% \end{equation} 
which is fixed at $-10$~dB and the transmit power of the legitimate transmitter is fixed at 30 dBm. All results are averaged over 500 random channel realizations.

% The performance is evaluated by 
% \begin{equation}
% \mathrm{SINR}(\boldsymbol{\theta})
% \triangleq
% \frac{P\left|H_0(\boldsymbol{\theta})\right|^2}
%      {P_J\sum_{k=1}^{K}\left|H_k(\boldsymbol{\theta})\right|^2 + \sigma_w^2},
% \end{equation}

% To demonstrate the effectiveness of the proposed method, several state-of-the-art benchmark schemes are considered for comparison.
% 1)\textbf{RM (RandomMax):} Specifically, $T$ candidate group phase shift vectors are randomly sampled from the discrete set $\mathcal{F}$, among which the one yielding the maximum observed SINR is selected.
% 2)\textbf{CSM (Conditional Sample Mean\cite{yao2023blind}):} After randomly generating $T$ group phase shift vectors and obtaining corresponding SINR values, each phase shift is selected by maximizing the conditional expection.
% 3)\textbf{Full CSI (Upper Bound):}
%     Assuming perfect CSI for all channels, problem (P1) is solved via manifold optimization \cite{boumal2014manopt}. This scheme serves as a performance upper bound for the proposed method.
% 4)\textbf{Full CSI (Discrete):}
%     Based on the Full CSI(upper-bound), the optimized reflection coefficients are further projected onto the discrete set. 

To demonstrate the effectiveness of the proposed method, several state-of-the-art benchmark schemes are considered for comparison:

\begin{itemize}%SS[leftmargin=0pt, label={}, labelsep=0.6em]
\item \textbf{RM(RandomMax):}  Specifically, $T$ candidate group phase shift vectors are randomly sampled from the discrete set $\mathcal{F}$ and the one that yields the maximum observed SINR is selected.
% , among which the one yielding the maximum observed SINR is selected.
\item \textbf{CSM(Conditional Sample Mean\cite{yao2023blind}):} 
After randomly generating $T$ group phase shift vectors and obtaining the corresponding SINR values, each phase shift is selected by maximizing the conditional expectation.
\item \textbf{Full CSI (continuous):}
Assuming perfect CSI for all channels, problem (P1) is solved using manifold optimization \cite{boumal2014manopt}. This approach serves as a performance upper bound for the proposed method.
\item \textbf{Full CSI:}
Based on the Full CSI (continuous), the optimized reflection coefficients are further projected onto the discrete set. 
\end{itemize}

% \begin{figure}[ht]
%     \centering
%     \includegraphics[width=0.95\linewidth]{PIC/figbit.eps} 
%     \caption{Average SINR versus the number of bits with $T=6000$.}
%     \label{bit_sweep}
% \end{figure}
Fig.~\ref{bit_sweep} illustrates the average SINR as a function of the number of quantization bits $B$ with a sufficiently large number of measurements $T=2000$. As expected, the full CSI (continuous) method provides an upper bound on the performance of all competing schemes. In contrast, the full-CSI approach experiences significant performance degradation when projecting onto the discrete phase set directly. The proposed method is inferior to the RM and CSM methods under 1-bit quantization but outperforms them by a significant margin as the number of quantization bits increases across different group sizes. The primary reason is that the gradient approximation in ZO-AdaMM is severely degraded under 1-bit quantization, whereas 2-bit resolution suffices to enable effective gradient estimation and realize substantial beamforming gains. Furthermore, the performance gap between different group sizes diminishes as the phase resolution increases to 5 bits, indicating that 2-bit phase shifters offer an optimal trade-off between hardware complexity and system performance for the considered scenario.
% For the proposed method, the SINR increases significantly from $B=1$ to $B=2$ for all tested group sizes, while further increasing the quantization bits beyond $B=2$ yields only marginal gains. Moreover, as $B$ increases, the performance gaps among different group sizes gradually shrink.
% This indicates that, with sufficient measurements, the proposed method can exploit most of the quantization gain using a small number of bits. 
% It indicates that, under sufficient measurements, the proposed method is able to exploit most of the quantization gain with a small number of quantization bits.
% , while the impact of grouping becomes less pronounced as $K$ grows.

% \begin{figure}[ht]
%     \centering
%     \includegraphics[width=0.95\linewidth]{PIC/fig1.eps}     
%     \caption{Average SINR versus the number of measurements $T$ for different group sizes.}
%     \label{meas_sweep}
% \end{figure}

Fig.~\ref{meas_sweep} illustrates the average SINR as a function of the number of the measurements $T$ for different group sizes. Both the proposed method and the CSM method achieve higher SINR performance as the number of measurements $T$ increases. Moreover, the proposed method significantly outperforms the RM and CSM methods by a remarkable margin. In addition, it can be observed that, under a limited number of measurements (e.g., $T\leq700$), a larger group size yields a higher SINR compared to a smaller group size, demonstrating a clear advantage in practical scenarios.
% it is observed that grouping accelerates convergence under limited measurements. Compared with the element-wise configuration, larger group sizes achieve higher SINR with fewer measurements, demonstrating improved sample efficiency. This advantage is particularly pronounced in the small-budget regime, where the reduction in effective search dimension enables faster exploration of the discrete phase space.
% As expected, the Manopt method continuous optimization provides an upper performance bound.

% \begin{figure}[ht]
%     \centering
%     \includegraphics[width=0.95\linewidth]{PIC/fig2.eps}   
%     \caption{Average SINR versus SIR with a fixed number of $T=600$.}
%     \label{sir_sweep}
% \end{figure}

Fig.~\ref{sir_sweep} depicts the average SINR versus SIR with a fixed number of measurements of $T=600$. The proposed method consistently outperforms RM and CSM across the entire SIR range for all tested group sizes. Moreover, the benefits of the proposed grouping method remain evident under moderate and strong interference conditions.
%, indicating that the proposed algorithm maintains robustness against interference when the measurements is fixed.

% \begin{figure}[htbp]
%     \centering
%     \includegraphics[width=0.8\linewidth]{PIC/fig3.eps} 
%     \caption{Average SINR versus the number of interference sources $K$ with $T=600$.}
%     \label{k_sweep}
% \end{figure}
% \textcolor{red}{Fig.~\ref{k_sweep} shows the average SINR as a function of the number of interference sources $K$ with a fixed measurements of $T=600$. When the number of interference sources is small, the proposed method achieves superior performance, benefiting from efficient grouping optimization. However, as $K$ increases, the SINR degrades for all schemes, and the performance gap narrows. }

\section{Conclusion}
% In this paper, we propose a Group-Based ZO-AdaMM algorithm, which can be further extended to discrete phase shifts via projection. Simulation results show that, compared with element-wise configuration, larger group sizes achieve higher SINR with fewer measurements, and the proposed method achieves faster, more sample-efficient convergence in the small-budget regime.

% In this paper, we propose a group-based ZO-AdaMM algorithm, which can be extended to discrete phase shifts via projection. Simulation results demonstrate that, compared with element-wise configurations, larger group sizes achieve higher SINR with fewer measurements, and the proposed method converges faster and exhibits improved sample efficiency in the small-budget regime.

In this paper, we developed a group-based ZO-AdaMM in which adjacent reflecting elements share a common phase shift through discrete phase projection. Simulation results demonstrated that the proposed approach achieves superior performance in terms of both SINR improvement and sample efficiency, making it well-suited for resource-constrained multi-RIS deployments.
% -------------------------------------------------------------------------
\bstctlcite{BSTcontrol}
\bibliographystyle{IEEEtran}
\bibliography{refs}

@string{ICC = {Proc. IEEE Int. Conf. Commun.}}

@article{10959084,
  author   = {Wang, Peilan and Fang, Jun and Wang, Bin and Li, Hongbin},
  journal  = {IEEE Transactions on Signal Processing},
  title    = {{Intelligent Reflecting Surface-Assisted Adaptive Beamforming for Blind Interference Suppression}},
  year     = {2025},
  volume   = {73},
  number   = {},
  pages    = {1744-1758},
  doi      = {10.1109/TSP.2025.3558965}
}

@article{liang2025survey,
  title     = {A Survey of Multiple Reconfigurable Intelligent Surface Aided Systems},
  author    = {Liang, Renjie and Liang, Junjie},
  journal   = {IEEE Access},
  year      = {2025},
  volume    = {13},
  number    = {},
  pages     = {194884-194895},
  publisher = {IEEE}
}

@article{zhao2025reconfigurable,
  title     = {{Reconfigurable Intelligent Surfaces for 6G: Engineering Challenges and the Road Ahead}},
  author    = {Zhao, Xianming and Jian, Mengnan and Chen, Yijian and Zhao, Yajun and Mu, Lin},
  journal   = {Intelligent and Converged Networks},
  volume    = {6},
  number    = {1},
  pages     = {53--81},
  year      = {2025},
  publisher = {TUP}
}

@article{wang2025low,
  title     = {{Low-Rank Covariance Matrix Recovery From Rank-One Measurements: An Analytical Solution}},
  author    = {Wang, Peilan and Fang, Jun and Ma, Binyao and Wang, Bin and Leus, Geert},
  journal   = {IEEE Signal Processing Letters},
  year      = {2025},
  volume    = {32},
  number    = {},
  pages     = {2674-2678},
  publisher = {IEEE}
}

@ARTICLE{11353466,
  author={Wang, Peilan and Fang, Jun and Zeng, Xianlong and Wang, Bin and Chen, Zhi and Eldar, Yonina C.},
  journal={IEEE Wireless Communications}, 
  title={{Derivative-Free Optimization-Empowered Wireless Channel Reconfiguration for 6G}}, 
  year={Early Access, 2026},
  pages={1-7},
}

@inproceedings{11310555,
  author    = {Ma, Binyao and Wang, Bin},
  booktitle = {2025 IEEE 102nd Vehicular Technology Conference (VTC2025-Fall)},
  title     = {{Blind Interference Suppression} for {IRS-Aided Communication Systems: A Derivative-free Optimization Approach}},
  year      = {2025},
  volume    = {},
  number    = {},
  pages     = {1-5},
  doi       = {10.1109/VTC2025-Fall65116.2025.11310555}
}

@article{basar2024reconfigurable,
  title    = {{Reconfigurable Intelligent Surfaces} for {6G: Emerging Hardware Architectures, Applications, and Open Challenges}},
  author   = {Basar, Ertugrul and Alexandropoulos, George C. and Liu, Yuanwei and Wu, Qingqing and Jin, Shi and Yuen, Chau and Dobre, Octavia A. and Schober, Robert},
  journal  = {IEEE Vehicular Technology Magazine},
  volume   = {19},
  number   = {3},
  pages    = {27-47},
  doi      = {10.1109/MVT.2024.3415570}
}

@article{wu2024intelligent,
  title     = {{Intelligent Surfaces Empowered Wireless Network: Recent Advances} and the {Road} to {6G}},
  author    = {Wu, Qingqing and Zheng, Beixiong and You, Changsheng and Zhu, Lipeng and Shen, Kaiming and Shao, Xiaodan and Mei, Weidong and Di, Boya and Zhang, Hongliang and Basar, Ertugrul and others},
  journal   = {Proceedings of the IEEE},
  volume    = {112},
  number    = {7},
  pages     = {724--763},
  year      = {2024},
  publisher = {IEEE}
}

@inproceedings{yao2023blind,
  title        = {{Blind Beamforming} for {Multiple Intelligent Reflecting Surfaces}},
  author       = {Yao, Jiawei and Xu, Fan and Lai, Wenhai and Shen, Kaiming and Li, Xin and Chen, Xin and Luo, Zhi-Quan},
  booktitle    = {ICC 2023-IEEE International Conference on Communications},
  pages        = {871--876},
  year         = {2023},
  organization = {IEEE}
}

@inproceedings{wang2025creating,
  title        = {{Creating} an {Interference-Free Environment} via {Intelligent Reflecting Surface: A Blind Approach Without Knowledge} of {CSI}},
  author       = {Wang, Peilan and Ma, Binyao and Fang, Jun and Wang, Bin and Li, Hongbin},
  booktitle    = {IEEE Vehicular Technology Conference (VTC2025-Spring)},
  pages        = {1--5},
  year         = {2025},
  organization = {IEEE}
}

@article{wang2020intelligent,
  title     = {{Intelligent Reflecting Surface-Assisted Millimeter Wave Communications: Joint Active} and {Passive Precoding Design}},
  author    = {Wang, Peilan and Fang, Jun and Yuan, Xiaojun and Chen, Zhi and Li, Hongbin},
  journal   = {IEEE Transactions on Vehicular Technology},
  volume    = {69},
  number    = {12},
  pages     = {14960--14973},
  year      = {2020},
  publisher = {IEEE}
}

@article{yan2025power,
  title     = {{Power Measurement Enabled Channel Autocorrelation Matrix Estimation} for {IRS-Assisted Wireless Communication}},
  author    = {Yan, Ge and Zhu, Lipeng and Zhang, Rui},
  journal   = {IEEE Trans. Wireless Commun.},
  publisher = {IEEE},
  year      = {2025},
  volume    = {24},
  number    = {3},
  pages     = {1832-1848}
}

@article{huang2019reconfigurable,
  title     = {{Reconfigurable Intelligent Surfaces for Energy Efficiency in Wireless Communication}},
  author    = {Huang, Chongwen and Zappone, Alessio and Alexandropoulos, George C and Debbah, M{\'e}rouane and Yuen, Chau},
  journal   = {IEEE Transactions on Wireless Communications},
  volume    = {18},
  number    = {8},
  pages     = {4157--4170},
  year      = {2019},
  publisher = {IEEE}
}

@article{boumal2014manopt,
  title     = {{{Manopt}}, a {{Matlab Toolbox}} for {{Optimization on Manifolds}}},
  author    = {Boumal, Nicolas and Mishra, Bamdev and Absil, P-A and Sepulchre, Rodolphe},
  journal   = {The Journal of Machine Learning Research},
  volume    = {15},
  number    = {1},
  pages     = {1455--1459},
  year      = {2014},
  publisher = {JMLR. org}
}

@article{munoz2015algorithm,
  title     = {{{Algorithm Selection}} for {{Black-Box Continuous Optimization Problems: A Survey}} on {{Methods}} and {{Challenges}}},
  author    = {Mu{\~n}oz, Mario A and Sun, Yuan and Kirley, Michael and Halgamuge, Saman K},
  journal   = {Information Sciences},
  volume    = {317},
  pages     = {224--245},
  year      = {2015},
  publisher = {Elsevier}
}

@article{chen2019zo,
  title   = {{ZO-AdaMM: Zeroth-Order Adaptive Momentum Method} for {Black-Box Optimization}},
  author  = {Chen, Xiangyi and Liu, Sijia and Xu, Kaidi and Li, Xingguo and Lin, Xue and Hong, Mingyi and Cox, David},
  journal = {Advances in Neural Information Processing Systems},
  volume  = {32},
  year    = {2019}
}

\end{document}